\documentclass[prl,twocolumn,showpacs,amsmath,amssymb]{revtex4-1}
\usepackage{subfigure}
\usepackage{graphicx}
\usepackage{graphicx}
\usepackage{array}
\usepackage{xcolor}

\usepackage{lineno}
\usepackage{defs_LamcNbar}
\usepackage{hyperref}

\begin{document}
\normalsize
\parskip=5pt plus 1pt minus 1pt

\title{Measurement of the absolute branching fraction of the inclusive decay $\bar{\Lambda}_{c}^{-} \to \bar{n} + X$}

\author{
\begin{small}
\begin{center}
M.~Ablikim$^{1}$, M.~N.~Achasov$^{11,b}$, P.~Adlarson$^{70}$, M.~Albrecht$^{4}$, R.~Aliberti$^{31}$, A.~Amoroso$^{69A,69C}$, M.~R.~An$^{35}$, Q.~An$^{66,53}$, X.~H.~Bai$^{61}$, Y.~Bai$^{52}$
% ...
\\
\vspace{0.2cm}
(BESIII Collaboration)\\
\vspace{0.2cm} {\it
$^{1}$ Institute of High Energy Physics, Beijing 100049, People's Republic of China\\
% ...
$^{4}$ Bochum Ruhr-University, D-44780 Bochum, Germany\\
% ...
$^{11}$ G.I. Budker Institute of Nuclear Physics SB RAS (BINP), Novosibirsk 630090, Russia\\
% ...
$^{31}$ Johannes Gutenberg University of Mainz, Johann-Joachim-Becher-Weg 45, D-55099 Mainz, Germany\\
% ...
$^{35}$ Liaoning Normal University, Dalian 116029, People's Republic of China\\
% ...
$^{52}$ Southeast University, Nanjing 211100, People's Republic of China\\
$^{53}$ State Key Laboratory of Particle Detection and Electronics, Beijing 100049, Hefei 230026, People's Republic of China\\
% ...
$^{61}$ University of Jinan, Jinan 250022, People's Republic of China\\
% ...
$^{66}$ University of Science and Technology of China, Hefei 230026, People's Republic of China\\
% ...
$^{69}$ University of Turin and INFN, (A)University of Turin, I-10125, Turin, Italy; (B)University of Eastern Piedmont, I-15121, Alessandria, Italy; (C)INFN, I-10125, Turin, Italy\\
$^{70}$ Uppsala University, Box 516, SE-75120 Uppsala, Sweden\\
% ...
$^{b}$ Also at the Novosibirsk State University, Novosibirsk, 630090, Russia\\
% ...
}
\end{center}
\vspace{0.4cm}
\end{small}
}

\begin{abstract}

Based on $\ee$ collision data corresponding to an integrated luminosity of 4.5 $\invfb$
collected at the center-of-mass energies between 4.600 and 4.699~$\gev$ with the BESIII detector at BEPCII,
the absolute branching fraction of the inclusive decay $\LamCNXB$,
where $X$ refers to any possible final state particles, is measured. 
The absolute branching fraction is determined to be $\mathcal{B}(\LamCNXB) = (32.4 \pm 0.7 \pm 1.5)\% $,
where the first uncertainty is statistical and the second systematic. 
Assuming $CP$ symmetry, the measurement indicates that about one-fourth of 
$\LamC$ ($\bar{\Lambda}_{c}^-$) decay modes with a neutron (an anti-neutron) in the final state have not been observed.
\end{abstract}

\maketitle

The $\LamC$ is the lightest charmed baryon, and the measurement of the properties of $\LamC$ provides key
input for studying heavier charmed baryons~\cite{Sheng:2018dc} and bottom baryons~\cite{Rupak:2016bb, Detmold:2015bb},
as well as understanding the dynamics of light quarks in the environment with a heavy quark~\cite{Cheng:2021qpd}.
However, there is no reliable phenomenological model calculation describing the
complicated physics of charmed baryon decays. 
Therefore, comprehensive and precise experimental studies of the $\LamC$ decays are highly desirable.

Experimentally, since the discovery of the $\LamC$ baryon in 1979~\cite{Abrams:1980Lc}, 
which eventually decays to a proton or a neutron,
its decays with a proton in the final state have been studied extensively.
However, information about decays with a neutron in the final state is sparse.
Recently, the BESIII collaboration measured the absolute branching fraction of 
decay $\LamCNPi$ to be $(6.6 \pm 1.2 \pm 0.4) \times 10^{-4}$~\cite{Ablikim:2022SCS},
where the double-tag (DT) approach~\cite{MarkIII:DT} is used, and the neutrons are
treated as missing particles and inferred under the laws of conservation of energy and momentum. 
This is the first-time measurement of singly Cabibbo suppressed mode involving a neutron directly 
in the final state in the $\LamC$ decays. 
Up to now, there are still very few measurements that directly observed neutron signals
in the $\LamC$  decays~\cite{Ablikim:2022SCS,Ablikim:2016mcr,Ablikim:2022nXpi},
including the decays $\LamC\to \Sigma^-2\pi^+$ and $\Sigma^-\pi^02 \pi^+$~\cite{Ablikim:sigmaXpi} 
where the $\Sigma^-$ is reconstructed with its dominant decay mode $\Sigma^-\to n\pi^-$. 
All the measurements implicitly include charge-conjugate modes.
Combing all the known exclusive decays of $\LamC$ summarized by the
Particle Data Group (PDG)~\cite{PDG:2020}, the total branching fraction of the
decays with a proton or a neutron in the final state is about 44\% or 25\%, respectively, which
include both the direct decay channels of $\LamC$ and the decays from intermediate particles,
$i.e.$  $\Lambda$, $\Sigma$, and $\Xi$.
There are still lots of unknown decay channels of $\LamC$ baryon to be explored experimentally.

The inclusive decay $\LamCNX$, where $X$ refers to any possible particle system,
has not yet been studied experimentally, due to the
difficulty in discriminating neutron signals from neutral noises.
In 1992, Ref.~\cite{Crawford:1992} estimated the inclusive branching fractions
of both $\LamC\to p+X$ and $\LamCNX$ to be $(50\pm16)\%$,
inferred from the known exclusive $B$-meson decays
and the fact that all $\LamC$ must decay to either proton or neutron. 
High precision measurements on the inclusive decays of $\LamC$ are crucial to point out the direction 
of searches for unknown channels. Furthermore, the results of inclusive decays will provide
direct information whether there exists a significant difference between the decays of  $\LamC$ with a proton
and a neutron in the final state. The investigation of the isospin symmetry between them 
is important input to theoretical estimation of the lifetime of the charmed baryon $\LamC$.

Comparing to the neutron, an anti-neutron has larger  energy deposition in an electromagnetic calorimeter (EMC)
due to its annihilation reaction with materials, which allows for good discrimination against
the contamination from the electromagnetic showers of photon. Hence, our measurement is 
conducted with the anti-particle decay $\LamCNXB$, which is supposed to yield the same result
as the $\LamCNX$ if the $\CP$ violation effect is ignored.

In this Letter, taking advantage of the excellent BESIII detector performance and of the $\LCpair$ production
just above the mass threshold,
the first measurement of absolute branching fraction of the $\LamCNXB$ decay 
is reported using $\ee$ collision data collected with the BESIII detector at seven 
center-of-mass (c.m.) energies between 4.600 and {\spaceskip=0.2em\relax 4.699 $\gev$}, 
corresponding to an integrated luminosity of 4.5 $\invfb$.
The integrated luminosities at these c.m. energies~\cite{BESIII:Lumi1, BESIII:Lumi2}
are summarized in \tablename~\ref{tab:yield-st}.

A detailed description of the design and performance of the BESIII detector can be found in Ref.~\cite{Ablikim:2009aa}.
Simulated samples are produced with a {\sc geant4}-based~\cite{Agostinelli:2002hh} Monte Carlo (MC) toolkit, 
which includes the geometric description of the BESIII detector. 
The signal MC samples of $\ee\to\LCpair$, with $\LamC$ decaying into the specific tag mode $\LamC \to \pkpi$
and $\LamCB$ going to any possible processes containing the already 
measured~\cite{Ablikim:2022SCS,Ablikim:2016mcr,Ablikim:2022nXpi,Ablikim:sigmaXpi,PDG:2020} and 
predicted~\cite{Geng:2019three} ones with an $\bar{n}$ in the final state, 
are used to determine the detection efficiencies.
They are generated for each individual c.m.~energy with the generator {\sc kkmc}~\cite{Jadach:2000ir}
by incorporating initial-state radiation (ISR) effects and the beam energy spread. 
The $\bar{n}$ candidates include the ones both from the interaction point (IP) and
from intermediate particles, $i.e.$ $\Lambda$, $\Sigma$ and $\Xi$.
The inclusive MC samples, which consist of 
$\LCpair$, charmed meson $D_{(s)}^{(\ast)}$ pair production,
ISR return to the charmonium(-like) $\psi$ states at lower masses,
and continuum processes $e^{+}e^{-}\rightarrow q\bar{q}$ ($q=u,d,s$), are generated to survey potential backgrounds.
Particle decays are modeled with
{\sc evtgen}~\cite{Lange:2001uf, Ping:2008zz} using branching
fractions taken from the PDG~\cite{PDG:2020}, when available, or otherwise estimated with
{\sc lundcharm}~\cite{Chen:2000tv, YANG:2014}.  Final state radiation
from charged final state particles is incorporated using {\sc
photos}~\cite{Richter-Was:1992hxq}.

The DT approach~\cite{MarkIII:DT} is implemented to measure the absolute branching fraction of $\LamCNXB$. 
Taking advantage of a large branching fraction and a high signal-to-background ratio, $\LamC$ baryons are reconstructed
in the $\LamC \to \pkpi$ decay mode, and are referred to as the single-tag (ST) candidates.
Events in which the signal decay $\LamCNXB$ is reconstructed in the system recoiling 
against the $\LamC$ candidates of the ST sample are denoted as the DT candidates. 

Charged tracks detected in the helium-based multilayer drift
chamber (MDC) are required to be within a polar angle ($\theta$) 
range of $|\!\cos\theta| < 0.93$, where $\theta$ is defined with respect to 
the $z$ axis, which is the symmetry axis of the MDC.
Their distances of the closest approach to the IP must be less than 10~cm along the $z$ axis, and less than 1~cm in the transverse plane.
The particle identification (PID) is implemented 
by combining measurements of the ionization energy loss ($\mathrm{d}E/\mathrm{d}x$) in the MDC and the flight time in the time-of-flight system, 
and to each charged track a particle type of pion, kaon, or proton is assigned, 
according to which assignment has the highest probability. 

The ST $\LamC$ candidates are identified using the
beam constrained mass $M_\mathrm{BC} = \sqrt{\Ebeam^2/c^4 - | \pLC |^2/c^2}$
and energy difference $\dE = E_{\LamC} - \Ebeam $, where $\Ebeam$ is the beam energy,
and $E_{\LamC}$ ($\pLC$) is the energy (momentum) of the $\LamC$ candidates in the c.m. frame.
The $\LamC$ candidates are required to satisfy the requirement $\dE \in (-34,~20)$~MeV. 
The asymmetric interval takes into account the effects of ISR and
corresponds to three times the resolution around the peak.
If there is more than one $\pkpi$ combination satisfying the above requirements, 
the one with the minimum $|\dE|$ is kept.

\begin{figure}[!htp]
    \begin{center}
        \includegraphics[width=0.45\textwidth, trim=5 0 0 0, clip]{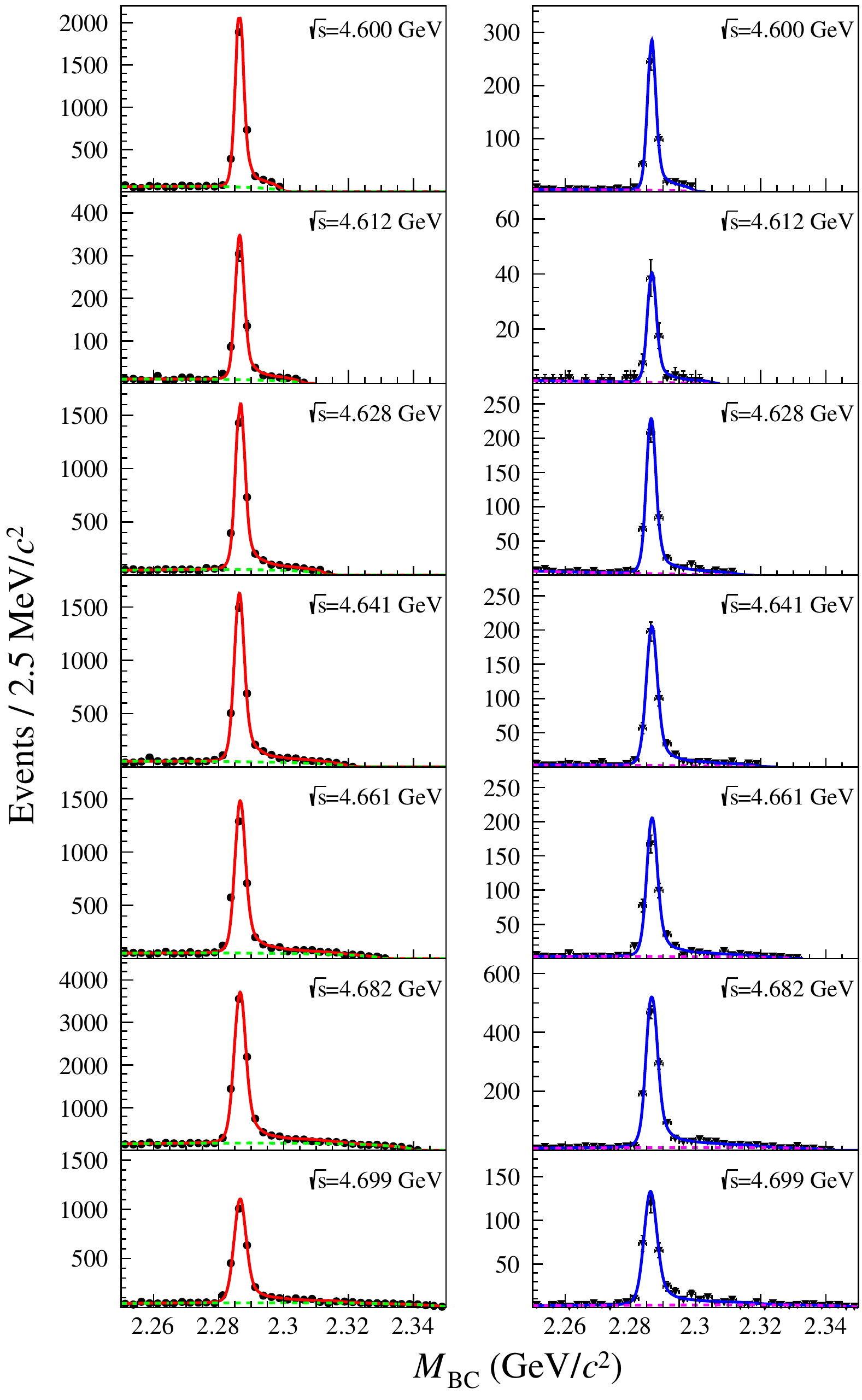}
    \end{center}
    \caption{
      The $\mBC$ distributions of $\LamC$ at seven c.m.~energies, and the data distributions
      are described by points (left column) or triangles (right column) with error bars. 
      The seven figures on the left column represent the results after the ST selections,
      and those on the right are obtained with both ST and $\bar{n}$ selections. 
      The (red) solid curves indicate the fit results and the (green) dashed curves describe the background shapes after the ST selections.
      The (blue) solid curves indicate the fit results and the (pink) dashed curves describe the background shapes after applying the $\bar{n}$ selections.
     }
     \label{fig:single-tag}
\end{figure}

The $\mBC$ distributions of candidate events for the ST mode
with data samples at different c.m.~energies are illustrated in \figurename~\ref{fig:single-tag},
where clear $\LamC$ signals are observed.
No peaking backgrounds are found with the investigation of the inclusive MC samples.
To obtain the ST yields,
unbinned maximum likelihood fits on these $\mBC$ distributions are performed,
where the signal is modeled with the MC-simulated distribution
convolved with a Gaussian function taking into account the resolution difference between data and MC simulation,
and the background distribution is described by an ARGUS function~\cite{ARGUS:1990hfq}
with the truncation parameter fixed to the corresponding $\Ebeam$.
The candidates within $\mBC \in (2.275,2.31)$~GeV/$c^2$ are retained for further analysis,
and the signal yields for the data samples at different c.m.~energies are summarized
in \tablename~\ref{tab:yield-st}.
The sum of ST yields for all data samples is $24,577 \pm 179$, where the uncertainty is statistical.

\begin{table}[!htbp]
  \begin{center}
  \caption{The integrated luminosity ($\mathcal{L}_{\rm int}$), ST yields, and the detection efficiencies of the ST and DT selections for the data samples at seven c.m.~energies.
            The uncertainties are statistical only. }
  \renewcommand\arraystretch{1.2}
  \resizebox{1.0\columnwidth}{!}{
    \begin{tabular}{ l p{1cm}<{\raggedleft} @{ $\pm$ } p{0.5cm}<{\raggedright}  p{1cm}<{\raggedleft} @{ $\pm$ } p{0.5cm}<{\raggedright} c c}
      \hline
      \hline
            $\sqrt{s}$ (GeV)   &   \multicolumn{2}{c}{$\mathcal L_{\rm int}$ (\ipb)}  &   \multicolumn{2}{c}{$N_{i}^{\mathrm{ST}}$}  & \makebox[0.1\textwidth][c]{$\epsilon_{i}^{\mathrm{ST}}$(\%)} & \makebox[0.1\textwidth][c]{$\epsilon_{i}^{\mathrm{DT}}$ (\%)} \\
      \hline

            4.600  &  586.9  & 0.1   &  3266  & 62   &  51.0 $\pm$ 0.2  & 19.1 $\pm$ 0.1 \\
            4.612  &  103.8  & 0.1   &  587   & 28   &  50.2 $\pm$ 0.2  & 19.2 $\pm$ 0.1 \\
            4.628  &  521.5  & 0.1   &  2967  & 64   &  49.5 $\pm$ 0.2  & 19.1 $\pm$ 0.1 \\
            4.641  &  552.4  & 0.1   &  3201  & 66   &  49.0 $\pm$ 0.2  & 18.9 $\pm$ 0.1 \\
            4.661  &  529.6  & 0.1   &  3080  & 63   &  48.0 $\pm$ 0.2  & 18.5 $\pm$ 0.1 \\
            4.682  &  1669.3 & 0.2   &  8863  & 107  &  47.3 $\pm$ 0.2  & 18.2 $\pm$ 0.1 \\
            4.699  &  536.5  & 0.1   &  2613  & 59   &  46.4 $\pm$ 0.2  & 17.8 $\pm$ 0.1 \\
      \hline\hline
    \end{tabular}
     }
          \label{tab:yield-st}
  \end{center}
\end{table}

\begin{figure*}[!htp]
    \begin{center}
            \subfigure[]{\includegraphics[scale=0.285]{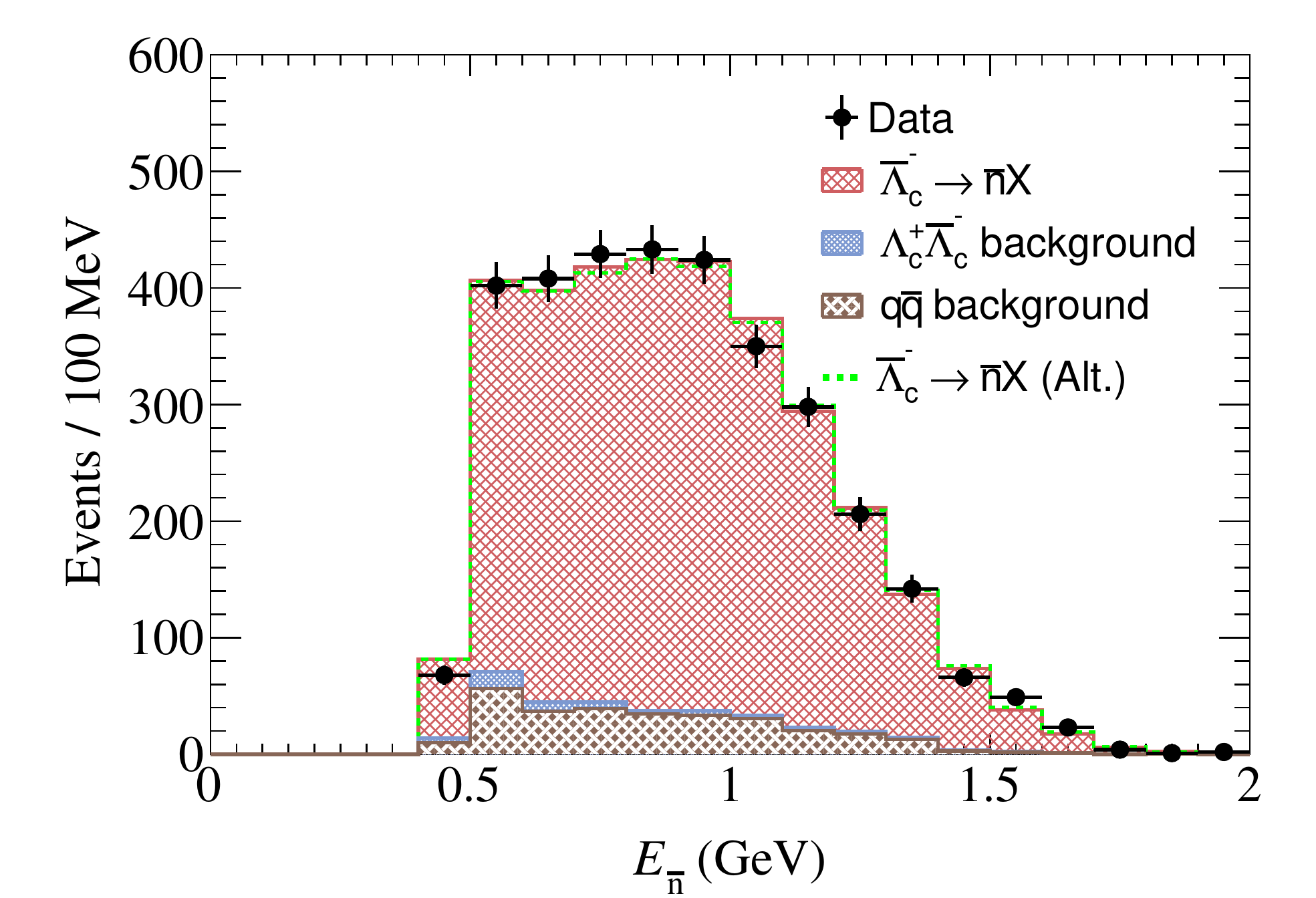}\label{fig:DT_signal_e}}
            \subfigure[]{ \includegraphics[scale=0.285]{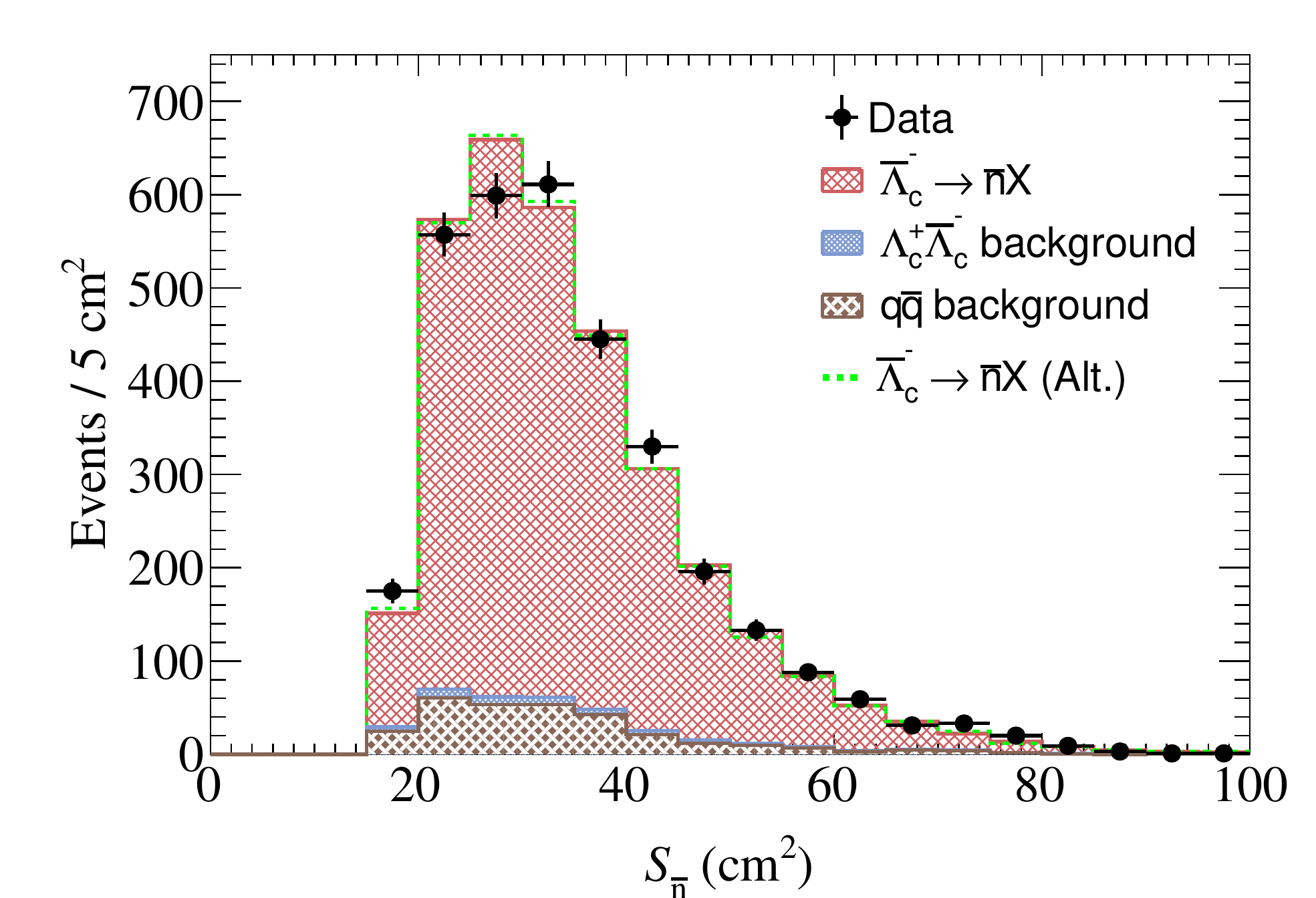}\label{fig:DT_signal_s}}      
            \subfigure[]  {  \includegraphics[scale=0.285]{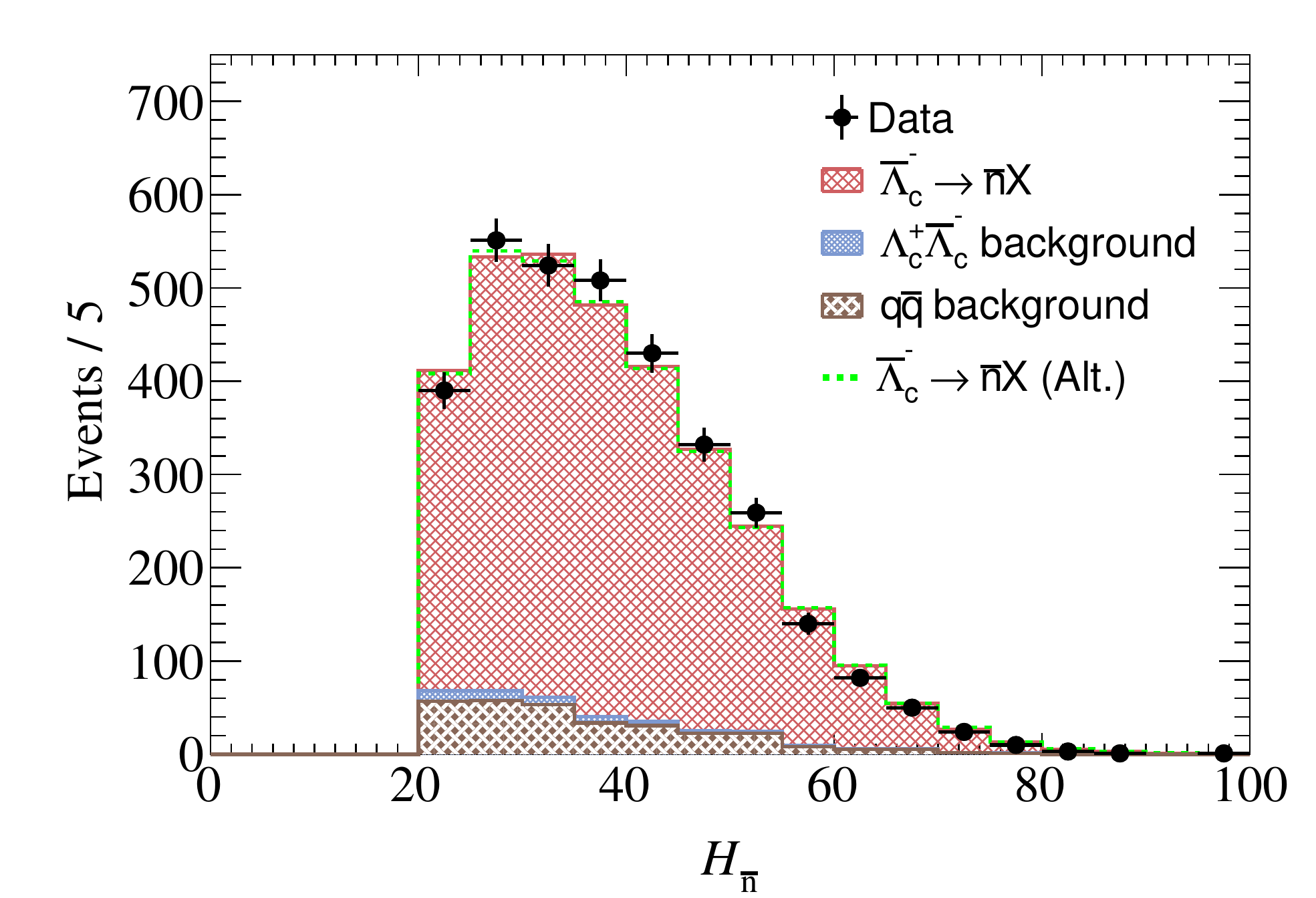}\label{fig:DT_signal_h} }
	\end{center}
    \caption{ The stacked distribution of $E_{\bar{n}}$ (a),  $S_{\bar{n}}$ (b) and $H_{\bar{n}}$ (c)
                   for the accepted DT candidates in the region 
                   $\mBC \in (2.275,2.31)$~GeV/$c^2$ from the combined seven data samples.                    
                  The black points with error bars are data. The red shaded histogram is the signal 
                  that is obtained with the data-driven method, and the green one describes the alternative signal
                  shape obtained with a Monte-Carlo sample with only the  observed decay modes.              
                  that is derived with only observed decay modes.
                  The blue and brown shaded histograms are the two background components,
                  where the $\LCpair$ is modeled with the inclusive MC sample of $\LamC\to p+X$
                  and the $q\bar{q}$ is estimated with events in the sideband region of $\mBC \in (2.20,2.26)$~GeV/$c^2$
                  and normalized to the region $\mBC \in (2.275,2.31)$~GeV/$c^2$. }
        \label{fig:DT_nbar}
\end{figure*}

The decay $\LamCNXB$ is searched for among the remaining tracks and showers recoiling against the ST $\LamC$ candidates.
Neutral showers are identified in the EMC. The deposited energy of each shower must be more than
25 MeV in the barrel region ($|\!\cos\theta| < 0.8$) and more than 50 MeV in the end cap region ($0.86<|\!\cos\theta| < 0.92$).
To suppress electronic noise and showers unrelated to the event, the difference between
the EMC time and the event start time is required to be within [0, 700] ns. 
The most energetic shower is taken as the $\bar{n}$ candidate. The angle between the charged track and shower is required to be greater than 20 degrees.
To discriminate the $\bar{n}$ shower from showers caused by photons and neutrons, three variables are used 
for further selection: the deposited energy ($E_{\bar{n}}$) in the EMC, 
the second moment of shower shape ($S_{\bar{n}} = \Sigma_{i}E_{i}r_{i}^{2}/\Sigma_{i}E_{i}$,
where $E_{i}$ is the energy deposited in
the $i^{\rm th}$ crystal of the shower and $r_{i}$ is the distance from the center of that crystals to the center of the shower),
and the number of hit crystals ($H_{\bar{n}}$) for the primary shower. 
The most energetic shower is required to have $E_{\bar{n}} > 0.48$ GeV, $H_{\bar{n}} > 20$, and $S_{\bar{n}} > 18$ cm$^2$.
To suppress contamination from the decays with a $\bar{p}$ particle in the final state, the
candidate events are further required to be without any 
tracks identified as $\bar{p}$ and having a distance of closest approach to the IP within 20~cm along the $z$ axis. 

In contrast to photons and electrons, the interaction of $\bar{n}$ with materials is very difficult to 
model, and there exists more than 10\% deviation in detection efficiency between data and MC simulation 
for the $\bar{n}$ induced clusters in the EMC.
To solve this issue, a model-independent data-driven method~\cite{Liang:2021nC} has been developed to 
simulate the detector response of the  $\bar{n}$ at BESIII. 
The detector response in data is investigated with a control sample of 16.2 million $\bar{n}$ candidates
selected in the process $J/\psi\to p \bar{n}\pi^-$ at $\sqrt{s}=3.097$~GeV~\cite{Ablikim:2022Jpsi}.
Firstly, the efficiency  of the requirements $E_{\bar{n}} > 0.48$ GeV, $H_{\bar{n}} > 20$, and $S_{\bar{n}} > 18$ cm$^2$
is derived in different finite bins ($\varepsilon_{\rm bin}$) of the two dimensional distribution of 
the momentum and polar angle $\cos\theta_{\bar{n}}$ of $\bar{n}$  by
comparing the yields of the $\bar{n}$ candidates in the control sample with and without
imposing the above requirements.
In the signal MC samples of the process $\LamCNXB$, each accepted event with these requirements
is determined,  if a random number, uniformly generated between 0 and 1,
is less than the value of $\varepsilon_{\rm bin}$ in the bin that the event belongs to. 
Then, the efficiency of the $\bar{n}$ selections is calculated by comparing the number of accepted events, summed over all bins,
with the total number of events at generator level. 
Secondly, the Probability Density Function and the Cumulative Distribution Function (CDF)
of the deposited energy $E_{\bar{n}}$, after applying the selection criteria,  are evaluated in these
different bins of momentum and $\cos\theta_{\bar{n}}$ with the control sample.
Then, the value of  $E_{\bar{n}}$ for each accepted event, in the signal MC samples, is sampled 
based on the CDF of $E_{\bar{n}}$ in the bin that 
the event belongs to.
After imposing all the selections mentioned above, 
the distribution of $E_{\bar{n}}$ for the accepted DT candidates 
from the combined data samples at seven c.m.~energies is shown in \figurename~\ref{fig:DT_nbar}, where
the data-driven method has been applied in the prediction of signal process $\LamCNXB$ and 
the simulated shape describes the data well.

The potential backgrounds can be classified into two categories: those directly originated
from continuum hadron production in the $\ee$ annihilation (denoted as $q\bar{q}$ background
hereafter) and those from the $\ee\to\LCpair$ events (denoted as $\LCpair$ background hereafter)
except for the signal of $\LamCNXB$. 
The resultant $E_{\bar{n}}$ distribution is depicted in \figurename~\ref{fig:DT_nbar}, 
where the events are selected in the region $\mBC \in (2.275,2.31)$~GeV/$c^2$. 
In \figurename~\ref{fig:DT_nbar}, the $q\bar{q}$ contamination, which is the major background component, 
is estimated with events in the sideband region $\mBC \in (2.20,2.26)$~GeV/$c^2$ and 
normalized to the region $\mBC \in (2.275,2.31)$~GeV/$c^2$.
The normalization factor is calculated with the event numbers in this two regions 
which are determined by integrating the ARGUS functions in the fitting to the ST $\mBC$ distributions.
The $\LCpair$ background is modeled with the inclusive MC sample of $\bar{\Lambda}_{c}^-\to \bar{p}+X$.

The yield of signal $\LamCNXB$ is obtained by performing unbinned maximum-likelihood fits on the 
$\mBC$ distributions of ST $\LamC$ after applying the $\bar{n}$ selections. 
The procedure is similar to the one used to obtain the ST yields.
The fitting curves for data samples at different c.m.~energies are illustrated in \figurename~\ref{fig:single-tag},
and the signal yields are obtained within $\mBC \in (2.275,2.31)$~GeV/$c^2$.
The $\LCpair$ contamination has the same shape as the signal process due to the undetected $\bar{p}$ tracks
and mis-identified $\bar{n}$ showers, and it is estimated with the inclusive MC samples and subtracted from 
observed signal yields. 
The fitting results and $\LCpair$ background are summarized in \tablename~\ref{tab:signalyield_fitmBC}.

\begin{table}[!htbp]
  \begin{center}
  \caption{Yields of the fitting results and the corresponding background estimation for the data samples at different c.m. energies.
            The uncertainty is statistical only. }
  \renewcommand\arraystretch{1.2}
    \begin{tabular}{ l p{1cm}<{\raggedleft} @{ $\pm$ } p{1cm}<{\raggedright} p{1cm}<{\raggedleft} @{ $\pm$ } p{1cm}<{\raggedright} }

      \hline
      \hline
           $\sqrt{s}$ (GeV)    &   \multicolumn{2}{c}{$N_{\mathrm{sig}}^{\mathrm{fit}}$}    &   \multicolumn{2}{c}{$N_{\mathrm{bkg-mc}}^{\LCpair}$}     \\
      \hline
         4.600      &        408 & 23          &      4.4 & 0.3              \\
         4.612      &        66  & 9             &      1.4 & 0.2              \\
         4.628      &        395 & 23          &      6.7 & 0.4              \\
         4.641      &        405 & 23          &      6.9 & 0.5              \\
         4.661      &        392 & 22          &      7.1 & 0.4              \\
         4.682      &        1135 & 36          &    20.5 & 0.6              \\
         4.699      &        304 & 19          &      5.8 & 0.4              \\
      \hline
         sum        &       3105 & 62          &     52.9 & 1.1              \\
      \hline\hline
    \end{tabular}

          \label{tab:signalyield_fitmBC}
  \end{center}
\end{table}

The branching fraction ($\mathcal{B}$) of decay $\LamCNXB$ is determined as
\begin{equation} \label{eq:br}
  \mathcal{B}=\frac{N_{\mathrm{sig}}^{\mathrm{DT}} - N_{\mathrm{bkg-mc}}^{\LCpair}} {\sum_{i} N_{i}^{\mathrm{ST}}\cdot (\epsilon_{i}^{\mathrm{DT}}/\epsilon_{i}^{\mathrm{ST}}) }
\nonumber
\end{equation}
where $N_{\mathrm{sig}}^{\mathrm{DT}}$ is the signal yield from the unbinned maximum-likelihood fit, and
$N_{\mathrm{bkg-mc}}^{\LCpair}$ is the estimated $\LCpair$ background from the inclusive MC samples.
The subscripts $i$ represents the data samples at different c.m. energies.
The parameters $N_{i}^{\mathrm{ST}}$, $\epsilon_{i}^{\mathrm{ST}}$ and $\epsilon_{i}^{\mathrm{DT}}$ are the ST yields, 
ST and DT efficiencies, respectively.
The ST and DT efficiencies are summarized in \tablename~\ref{tab:yield-st}, where
the efficiency of $\bar{n}$ selections is already corrected with the the data-driven method~\cite{Liang:2021nC}.
The branching fraction is determined to be $\mathcal{B}(\LamCNXB)=(32.4 \pm 0.7 \pm 1.5)\% $,
where the first uncertainty is statistical and the second systematic.

The systematic uncertainties for the branching fraction measurement include those associated with the ST yields, 
detection efficiencies of the ST $\LamC$ and  the DT selections.
As the DT technique is adopted, the systematic uncertainties associated with the ST detection efficiency cancel out. 

The uncertainty in the ST yields is 0.5\%, which arises from the statistical uncertainty and a systematic component 
coming from the fit to the $\mBC$ distribution. The uncertainty is evaluated by 
floating the truncation parameter of the ARGUS function
and changing the single gaussian function to a double gaussian function.
The uncertainty associated with the finite size of the signal MC samples is 0.3\%. 
The uncertainty arising from the signal modeling is 4.1\%, which combines two sources. 
The first is due to unknown processes in the MC production, which is investigated by 
generating alternative signal MC samples only with the known $\bar{n}$ processes in the PDG~\cite{PDG:2020}.
The second one is the imperfect simulation of the $E_{\bar n}$ distribution, which is estimated by comparing 
the difference in the detection efficiencies between the results with and without reweighting the 
MC-simulated $E_{\bar n}$ distribution to data, 
where all the signal selection criteria in the analysis are applied except for the $E_{\bar{n}}$ requirement. 
For each case, the change of the signal efficiencies is taken as the systematic uncertainty. 
The uncertainty in the fit strategy of extracting the signal yields is assigned to be 0.4\%, which is estimated by
floating the truncation parameter of the ARGUS function and changing the single gaussian function to a double gaussian function.
The uncertainty arising from $\LCpair$ background estimation is studied by 
generating alternative inclusive MC samples only with the known processes in the PDG~\cite{PDG:2020},
and comparing the background yields between the nominal and alternative MC samples.
The difference of signal yields after subtracting the estimated $\LCpair$ background,
1.0\%, is assigned as the corresponding uncertainty.
The uncertainty due to $\bar{n}$ selections is assigned to be 2.0\%, as explained in Ref.~\cite{Liang:2021nC}. 
All other uncertainties are negligible. 
Assuming that all the systematic uncertainties are uncorrelated, the total uncertainty is then taken to be 
the quadratic sum of the individual values, which is 4.7\%.

In summary, the first measurement of the absolute branching fraction of the inclusive decay $\LamCNXB$ is reported using
4.5~$\invfb$ $\ee$ collision data collected at seven
c.m.~energies between 4.600 and 4.699 GeV with the BESIII detector.
The absolute branching fraction is determined to be $\mathcal{B}(\LamCNXB) = (32.4 \pm 0.7 \pm 1.5)\% $,
where the first uncertainty is statistical and the second systematic.
Neglecting the effect of $\CP$ violation,
the inclusive decay $\LamCNX$ should have the same value as $\LamCNXB$.
The measurement significantly improves the precision up to 5\%, 
comparing to the previous result of this inclusive decay, ($50 \pm 16$)\%, 
inferred from the $B$-meson decays~\cite{Crawford:1992}. 
The branching fraction of sum over all the known exclusive decays with a neutron in the final state  
is about $(25.4\pm0.8)$\%~\cite{PDG:2020,Ablikim:2022SCS,Ablikim:2022nXpi},  
where the uncertainties of all the modes are treated without correlation.
It means that about one-fourth of the $\LamC$ decays with a neutron in the final state remain to be explored
in experiments. 

The BESIII collaboration thanks the staff of BEPCII and the IHEP computing center and the supercomputing center of
the University of Science and Technology of China (USTC) for their strong support. 
This work is supported in part by National Key R\&D Program of China under Contracts Nos. 2020YFA0406400, 2020YFA0406300; 
National Natural Science Foundation of China (NSFC) under Contracts Nos. 11635010, 11735014, 11835012, 11935015, 11935016, 
11935018, 11961141012, 12022510, 12025502, 12035009, 12035013, 12192260, 12192261, 12192262, 12192263, 12192264, 12192265, 12005311; 
the Fundamental Research Funds for the Central Universities, Sun Yat-sen University, University of Science and Technology of China;
100 Talents Program of Sun Yat-sen University;
the Chinese Academy of Sciences (CAS) Large-Scale Scientific Facility Program; 
Joint Large-Scale Scientific Facility Funds of the NSFC and CAS under Contract No. U1832207; 
the CAS Center for Excellence in Particle Physics (CCEPP);
100 Talents Program of CAS; 
The Institute of Nuclear and Particle Physics (INPAC) and Shanghai Key Laboratory for Particle Physics and Cosmology; 
ERC under Contract No. 758462; 
European Union's Horizon 2020 research and innovation programme under Marie Sklodowska-Curie grant agreement under Contract No. 894790; 
German Research Foundation DFG under Contracts Nos. 443159800, Collaborative Research Center CRC 1044, GRK 2149; 
Istituto Nazionale di Fisica Nucleare, Italy; 
Ministry of Development of Turkey under Contract No. DPT2006K-120470; 
National Science and Technology fund; 
National Science Research and Innovation Fund (NSRF) via the Program Management Unit for Human Resources \& Institutional Development, 
Research and Innovation under Contract No. B16F640076; 
STFC (United Kingdom); 
Suranaree University of Technology (SUT), Thailand Science Research and Innovation (TSRI), 
and National Science Research and Innovation Fund (NSRF) under Contract No. 160355; 
The Royal Society, UK under Contracts Nos. DH140054, DH160214; 
The Swedish Research Council; 
U. S. Department of Energy under Contract No. DE-FG02-05ER41374

\end{document}